\documentclass[prd,showpacs]{revtex4-1}

\usepackage{amsmath}
\usepackage{amssymb}
\usepackage[dvips]{graphicx}
\usepackage{hyperref}

\begin{document}
\title{Rastall Cosmology and the $\Lambda$CDM Model}

\author{Carlos~E.~M.~Batista\footnote{cedumagalhaes22@hotmail.com}}
\affiliation{Departamento de F\'{\i}sica, Universidade Estadual de Feira de Santana, Feira de Santana, Brazil}
\author{Mahamadou~H.~Daouda\footnote{daoudah8@yahoo.fr}, J\'ulio~C.~Fabris\footnote{fabris@pq.cnpq.br}, Oliver~F.~Piattella\footnote{oliver.piattella@gmail.com}, Davi~C.~Rodrigues\footnote{davi.c.rodrigues@gmail.com}}
\affiliation{Departamento de F\'{\i}sica, Universidade Federal do Esp\'{\i}rito Santo, Vit\'oria, Brazil}

\begin{abstract}
Rastall's theory is based on the non-conservation of the energy-momentum tensor. We show that, in this theory, if we introduce a two-fluid model, one component representing vacuum energy whereas the other pressure-less matter (e.g. baryons plus cold dark matter), the cosmological scenario is the same as for the $\Lambda$CDM model, both at background and linear perturbation levels, except for one aspect: now dark energy may cluster. We speculate that this can lead to a possibility of distinguishing the models at the non-linear perturbation level.
\end{abstract}


\maketitle

\section{Introduction}

Since the formulation of General Relativity, about one hundred years ago, many alternative geometric theories have been proposed in order to explain gravitation phenomena (e.g. \cite{Cartan1922, Cartan1923, Brans1961, Rosen1971, Rastall1973, Rastall1976, Moffat:1994hv, Bekenstein:2004ne}). Some of these touch one important aspect of General Relativity: the conservation law. To our knowledge, the first non-conservative theory of gravity was the steady-state model \cite{Bondi1948, Hoyle1948}, following some ideas already presented in the end of the forties by Jordan \cite{Jordan:1949zz}. In the beginning of the seventies, Rastall proposed one new version of a non-conservative theory of gravity, following the remark that the conservation law ${T^{\mu\nu}}_{;\mu} = 0$ may not hold true in a curved space-time \cite{Rastall1973, Rastall1976}. Hence, he argued that new gravitational equations can be obtained considering a modification of the conservation law such that
\begin{eqnarray}
\label{mot1}
{T^{\mu\nu}}_{;\mu} = \kappa R^{;\nu},
\end{eqnarray}
where $T^{\mu\nu}$ is the energy-momentum tensor, $\kappa$ is a coupling constant and $R$ is the Ricci scalar curvature. Hence, in the weak field limit, the usual expressions are preserved. Since, generally, the Ricci scalar curvature is connected with the trace of the energy-momentum tensor, Eq.~(\ref{mot1}) can be re-written as
\begin{eqnarray}
{T^{\mu\nu}}_{;\mu} = \bar\kappa T^{;\nu}\;,
\end{eqnarray}
where $\bar\kappa$ is a new constant and $T$ is the trace of the energy-momentum tensor. It is curious to remark that the phenomenon of particle creation in cosmology \cite{Gibbons:1977mu, Parker:1971pt, Ford:1986sy} also leads to a violation of the classical conservation laws and, in this sense, Rastall's idea may be viewed as a kind of classical formulation of that quantum phenomenon, since the violation of the energy-momentum conservation is connected with the curvature.

More in detail, Rastall's modification to Einstein equations take the following form ($c = 1$ units):
\begin{eqnarray}
\label{eq1R} R_{\mu\nu} - \frac{1}{2}g_{\mu\nu}R &=& 8\pi G\left(T_{\mu\nu} - \frac{\gamma- 1}{2}g_{\mu\nu}T\right)\;,\\
\label{eq2R} {T^{\mu\nu}}_{;\mu} &=& \frac{\gamma - 1}{2}T^{;\nu}\;,
\end{eqnarray}
where $\gamma$ is a parameter (the choice $\gamma = 1$ restores General Relativity). Note that it seems possible to have a Lagrangian formulation from which the above equations are deduced \cite{Smalley1984}. Since for a radiative fluid $T = 0$, implying $R = 0$, we can expect that the cosmological evolution during the radiative phase is the same as in the standard cosmological scenario. At same time, a single fluid inflationary model, described by a cosmological constant, is the same as it would be in the General Relativity case. Hence, Rastall cosmologies may have an important departure from the standard cosmological model from the beginning of the matter dominated phase on \cite{Batista:2010nq, Fabris:2011rm, Fabris:2011wz, Capone:2009xm}.

\section{The Model}

In order to construct a Rastall cosmology which accounts for the matter dominated era and the present phase of acceleration of the universe, let us consider a two-fluid model.
The first component is a pressure-less matter (i.e. cold dark matter plus baryons) with density $\rho_m$, while the second one obeys the vacuum equation of state $p_x = - \rho_x$. A subscript (or superscript) $m$ shall denote quantities related to the matter component whereas a subscript (or superscript) $x$ shall refer to dark energy quantities.

Equations~\eqref{eq1R} and \eqref{eq2R} then become
\begin{eqnarray}
\label{EEqgen} R_{\mu\nu} - \frac{1}{2}g_{\mu\nu}R &=& 8\pi G\left[T^m_{\mu\nu} + T^x_{\mu\nu} - \frac{\gamma - 1}{2}g_{\mu\nu}(T^m + T^x)\right]\;,\\
\label{conseqgen} \left(T^{\mu\nu}_m + T^{\mu\nu}_x\right)_{;\mu} &=& \frac{\gamma - 1}{2}(T_m + T_x)^{;\nu}\;.
\end{eqnarray}
The latter equation fix the divergence of the total energy-momentum tensor, while extra assumptions are necessary in order to fix $(T^{\mu\nu}_m)_{;\mu}$ and $(T^{\mu\nu}_x)_{;\mu}$ (similarly to General Relativity, where the Bianchi identities only constrain the total energy-momentum tensor). In general, for arbitrary linear combinations of $T_m$ and $T_x$ such that  Eq.~(\ref{conseqgen}) is preserved, we introduce the arbitrary real parameters $\eta_x$ and $\eta_m$ as follows:
\begin{equation}
\label{conseqgenmx} T^{\mu \nu}_{m ;\mu} = \frac{\gamma - 1}{2}\left(\eta_m T_m + \eta_x T_x\right)^{;\nu}\;, \qquad T^{\mu \nu}_{x ;\mu} = \frac{\gamma - 1}{2}\left[(1 - \eta_m) T_m + (1 - \eta_x) T_x\right]^{;\nu}\;,
\end{equation}
Note that, for $\gamma = 1$, we recover the standard (General Relativity) case. 

Together with the flat Robertson-Walker metric,
\begin{equation}\label{RWmet}
ds^2 = dt^2 - a^2(t)\delta_{ij}dx^idx^j\;,
\end{equation} 
the Einstein and conservation equations \eqref{EEqgen}, \eqref{conseqgenmx} take the form
\begin{eqnarray}
\label{modFriedEq} H^2 &=& \frac{8\pi G}{3}\left[(3 - 2\gamma)\rho_x + \frac{3 - \gamma}{2}\rho_m\right]\;,\\
\dot \rho_m + 3 H \rho_m &=& \frac{\gamma - 1}{2}\left(\eta_m \dot \rho_m + 4 \eta_x \dot \rho_x\right)\;,\\
\dot \rho_x &=& \frac{\gamma - 1}2 \left[(1 - \eta_m) \dot \rho_m + 4 (1 - \eta_x) \dot \rho_x\right]\;.
\end{eqnarray}
Combining the second equation with the third, one obtains
\begin{equation}
\label{mattconsgen} \dot \rho_m + 3 H \rho_m = \frac{\gamma - 1}{2} \left[\frac{\eta_m + 2 (\gamma - 1) (\eta_x - \eta_m)}{1 + 2(\gamma - 1) (\eta_x-1)}\right] \dot \rho_m\;,\qquad
\dot \rho_x = \frac{\gamma -1}2 \frac{1 - \eta_m}{1 + 2 (\gamma-1)(\eta_x -1)} \dot \rho_m\;.
\end{equation}
Thus, we have the following solutions for the densities:
\begin{eqnarray}
\label{mattevoeta} \rho_m &=& \rho_{m0}a^{-3(1 + \omega_e)}\;, \qquad \omega_e \equiv \frac{\gamma - 1}{2} \frac{\eta_m + 2 (\gamma - 1) (\eta_x - \eta_m)}{1 + 2(\gamma - 1) (\eta_x-1) - \frac{\gamma - 1}{2}\left[\eta_m + 2 (\gamma - 1) (\eta_x - \eta_m)\right]}\;,\\
\label{xevoeta} \rho_x &=& \frac{\bar\rho}{3 - 2\gamma} + \frac{\gamma -1}2 \frac{1 - \eta_m}{1 + 2 (\gamma-1)(\eta_x -1)}\rho_m\;,
\end{eqnarray}
where $\bar\rho$ is an integration constant and we have introduced $\omega_e$ as an effective equation of state. Inserting Eq.~\eqref{xevoeta} in the modified Friedman equation \eqref{modFriedEq}, we obtain
\begin{equation}\label{FriedEqmod}
H^2 = \frac{8\pi G}{3}\left\{\bar\rho + \left[1 + \frac{(\gamma - 1)\left[\left(3 - 2\gamma\right)(1 - \eta_m) - 1 - 2(\gamma - 1)(\eta_x - 1)\right]}{2\left[1 + 2(\gamma - 1)(\eta_x - 1)\right]}\right]\rho_m\right\}\;,
\end{equation}
which would describe the same background evolution as in the $\Lambda$CDM model, with now $\bar\rho$ playing the role of an effective cosmological constant, if matter had not now a different evolution, depending on $\gamma$, $\eta_m$ and $\eta_x$.

\begin{figure}[ht]
\centering
\includegraphics[width=0.3\columnwidth]{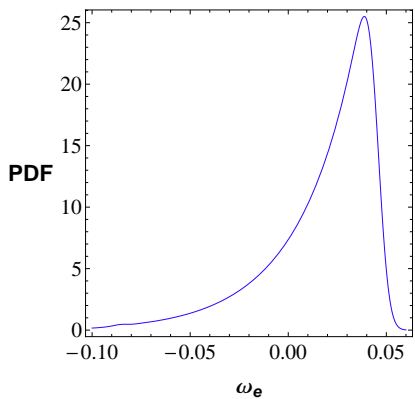}
\caption{Probability density function (PDF) of $\omega_e$.}
\label{Fig1PDF}
\end{figure}

Equation~(\ref{mattevoeta}) predicts a deviation, given by the parameter $\omega_e$, from the usual dependence of the matter component with respect to the scale factor. Indeed, in the $\Lambda$CDM model: $\omega_e = 0$. In order to constrain $\omega_e$, we use supernova type Ia data, Union2 sample \cite{Amanullah:2010vv}, $H(z)$ data \cite{Gaztanaga:2008xz, Stern:2009ep}, first CMB acoustic peak data given in terms of the $R$ factor \cite{Komatsu:2010fb}, and BAO data \cite{Komatsu:2010fb, Eisenstein:2005su}. We perform a bayesian statistical analysis (as done in \cite{Fabris:2011rm}) for the model defined by the cosmological term plus a fluid with $p_m = \omega_e\rho_m$. Three free parameters are consider: the matter density, the Hubble constant, and $\omega_e$. Marginalizing over the matter density and the Hubble constant, we find, at $2\sigma$: $\omega_e = 0.039^{+0.014}_{-0.080}$. The PDF for $\omega_e$ is displayed in Fig.~\ref{Fig1PDF}.

Another estimation can be obtained by using structure formation. Following \cite{Fabris:1998hr}, we see that the modification of the equation of state induced by Rastall's theory at the background level, from the point of view of the general relativity framework, i.e. $p_m = \omega_e\rho_m$, is maintained for linear perturbations, i.e. $\delta p_m \approx \omega_e \delta\rho_m$ (we use the approximate sign since in \cite{Fabris:1998hr} the model considered is a single-fluid one). The cold dark matter scenario implies $\omega_e = 0$, while the warm dark matter scenario could lead to $\omega_e \sim 0$ (we admit that the hot dark matter scenario is excluded by observation). Suggested by these remarks, we fix $\omega_e = 0$ from now on.

It is remarkable that even if $\gamma$ is significantly different from one, the cosmological background of Rastall's theory may lead to the same results of the $\Lambda$CDM model. To this end, we set $\omega_e = 0$, whose single $\gamma$-independent solution is 
\begin{equation}
 \eta_x = \eta_m = 0\;.
\end{equation}
The above values for $\eta_x$ and $\eta_m$ lead to a particular but interesting case of Rastall's theory. The above setting guarantees that the matter energy-momentum tensor is conserved in the usual sense, i.e. $T^{\mu \nu}_{m;\mu} = 0\;.$

In fact, with the above hypothesis, Eq.~\eqref{FriedEqmod} reduces to the corresponding one in the $\Lambda$CDM model, without any explicit dependence on $\gamma$.

\section{small perturbations}

We turn now to the fate of linear perturbations. We will work initially with the synchronous coordinate condition, since it leads directly to an expression for the density contrast of matter perturbations \cite{Padmanabhan1993, Ma:1995ey}.
It is more convenient, in this case, to use the field equations under the form,
\begin{equation}
R_{\mu\nu} = 8\pi G\left(T_{\mu\nu} - \frac{2 - \gamma}{2}g_{\mu\nu}T\right)\;,\qquad {T^{\mu\nu}}_{;\mu} = \frac{\gamma - 1}{2}T^{;\nu}\;.
\end{equation}
The synchronous coordinate condition implies that
\begin{equation}
\tilde g_{\mu\nu} = g_{\mu\nu} + h_{\mu\nu}\;, \quad h_{\mu0} = 0\;,
\end{equation}
where $\tilde g_{\mu\nu}$ is the total (perturbed) metric, $g_{\mu\nu}$ is the background metric and $h_{\mu\nu}$ are fluctuations over the latter.

The fluids are represented by the following energy-momentum tensors:
\begin{equation}
T_x^{\mu\nu} = (\rho_x + p_x)u_x^\mu u_x^\nu - p_xg^{\mu\nu}\;, \qquad T_m^{\mu\nu} =  (\rho_m + p_m)u_m^\mu u_m^\nu - p_mg^{\mu\nu}\;.
\end{equation}
Later, we shall impose $p_m = 0$ and $p_x = - \rho_x$.
The choice of the synchronous coordinate condition implies $\delta u^0 = 0$, but $\delta u^i$ is a dynamical variable. There are in fact two four-velocities, one associated to
the $x$ component, the other to the matter one. Using this formalism, one obtains the following set of equations for perturbations: 
\begin{eqnarray}
\label{pe1}
 \ddot h &+& 2\frac{\dot a}{a}\dot h = 8\pi G\rho_x[\gamma + 3(2 - \gamma)\omega_x]\delta_x + 8\pi G\rho_m\gamma\delta_m\;,\\
\label{deltamh}
 \dot\delta_m &=& \frac{\dot h}{2}\;,\\
\label{deltarhox}
\delta\dot\rho_x &+& (1 + \omega_x)\rho_x\biggr(\Theta - \frac{\dot h}{2}\biggl) + 3\frac{\dot a}{a}(1 + \omega_x)\delta\rho_x = 
\frac{\gamma - 1}{2}\left[(1 - 3\omega_x)\delta\dot\rho_x + \delta\dot\rho_m\right]\;,\\
\label{deltaui}
(1 + \omega_x)\dot\rho_x\delta u_x^i &+& (1 + \omega_x)\rho_x\delta\dot u_x^i + 5\frac{\dot a}{a}(1 + \omega_x)\rho_x\delta u_x^i +  \frac{\omega_x}{a^2}\partial^i\delta\rho_x
= -\frac{\gamma - 1}{2a^2}\left[(1 - 3\omega_x)\partial^i\delta\rho_x + \partial^i\delta\rho_m\right]\;.
\end{eqnarray}
In these expressions we have used the definitions
\begin{equation}
h \equiv \frac{h_{kk}}{a^2}\;, \quad \omega_x \equiv \frac{p_x}{\rho_x}\:, \quad \delta_x \equiv \frac{\delta\rho_x}{\rho_x}\;, \quad \delta_m \equiv \frac{\delta\rho_m}{\rho_m}\;, \quad \Theta \equiv \partial_i \delta u_x^i\;,
\end{equation}
where $\partial_i$ denotes derivative with respect to the co-moving spatial coordinates. Now, let us impose $\omega_x =  - 1$. We obtain, from Eqs.~(\ref{deltarhox}) and (\ref{deltaui}), that
\begin{equation}\label{r12}
\delta\dot\rho_x = \frac{\gamma - 1}{2}\left(4\delta\dot\rho_x + \delta\dot\rho_m\right)\;,\qquad
-\frac{\partial^i\delta\rho_x}{a^2} = \frac{1 - \gamma}{2a^2}\left(4\partial^i\delta\rho_x + \partial^i\delta\rho_m\right)\;,
\end{equation}
and both these equations lead to the relation
\begin{eqnarray}
\label{fpr1}
\delta\rho_x = \frac{\gamma - 1}{2(3 - 2\gamma)}\delta\rho_m\;.
\end{eqnarray}
Using now Eq.~(\ref{deltamh}) one can rewrite Eq.~(\ref{pe1}) as
\begin{equation}
\ddot\delta_m + 2\frac{\dot a}{a}\delta_m - 4\pi G\rho_m\delta_m = 0\;.
\end{equation}
But this is the same equation for matter perturbation as in the $\Lambda$CDM model! Actually, there is a difference: now there are perturbations in the dark energy term $\rho_x$. In fact, using Eqs.~(\ref{mattevoeta}) and (\ref{xevoeta}) with $\eta_m = \eta_x = 0$ together with (\ref{fpr1}), we find
\begin{equation}\label{deltaxdeltamrel}
\delta_x = \frac{\gamma - 1}{2}\frac{\delta_m}{\frac{\bar\rho}{\rho_{m0}}a^3 + \frac{\gamma - 1}{2}}\;.
\end{equation}
In the remote future (i.e. for $a \rightarrow \infty$), $\delta_x$ must become negligible, and a complete equivalence between the Rastall and the $\Lambda$CDM model is expected both for the background expansion and the linear perturbations evolution. On the other hand, in the remote past (i.e. for $a \rightarrow 0$), we have $\delta_x \sim \delta_m$ and therefore dark energy may in principle cluster. This may have consequences for the structure formation process because we should expect an amount of clustered dark energy to be present in virialized systems, like halos of galaxies and clusters of galaxies. However, as we have shown, such consequences seem not to appear for linear perturbations.

\subsection{Newtonian gauge}

A possibility to discriminate between the two models could be a different integrated Sachs-Wolfe effect signal \cite{Sachs1967}. To see if this is the case, we adopt now the conformal-Newtonian one. Consider the following perturbations to the Robertson-Walker metric expressed in the conformal time $\eta\;:$
\begin{equation}\label{metpert}
 ds^2 = a^2(\eta)(1 + 2\Phi)d\eta^2 - a^2(\eta)(1 - 2\Phi)\delta_{ij}dx^idx^j\;,
\end{equation}
where $\Phi(\eta, x^i)$ is the gravitational potential. For each fluid component we calculate the perturbations to the continuity equation and to the Euler one. Following \cite{Mukhanov2005}, for the case of matter we have
\begin{equation}\label{consmatt}
 \delta\rho_m' + 3\frac{a'}{a}\delta\rho_m + a\rho_m\partial_i\delta u_m^i - 3\rho_m\Phi' = 0\;, \qquad \frac{1}{a^4}\left(a^5\rho_m\delta u_m^i\right)^\prime + \rho_m\partial^i\Phi = 0\;,
\end{equation}
where the prime denotes derivation with respect to the conformal time. For the $x$ component, we have the correction proportional to $T_{,\mu}$ and the equations are
\begin{eqnarray}\label{contpertx}
 \delta\rho_x' + 3\frac{a'}{a}(\delta p_x + \delta\rho_x) -3\rho_x(1 + w_x)\Phi' + a\rho_x(1 + w_x)\partial_i\delta u_x^i &=& \frac{\gamma - 1}{2}\left[\delta\rho_m' + \delta\rho_x' - 3\delta p_x'\right]\;,\\ \label{eulerpertx}
 \frac{1}{a^4}\left[a^5\rho_x(1 + w_x)\delta u_x^i\right]^\prime + \partial^i\delta p_x + \rho_x(1 + w_x)\partial^i\Phi &=& -\frac{\gamma - 1}{2}\partial^i\left[\delta\rho_m + \delta\rho_x - 3\delta p_x\right]\;.
\end{eqnarray}
For the Einstein equations, we have
\begin{eqnarray}\label{E00}
 \Delta\Phi - 3\frac{a'}{a}\left(\frac{a'}{a}\Phi + \Phi'\right) = 4\pi G a^2\left[\delta\rho_m + \delta\rho_x - \frac{\gamma - 1}{2}\left(\delta\rho_m + \delta\rho_x - 3\delta p_x\right)\right]\;,\\
\label{Eij} \Phi'' + 3\frac{a'}{a}\Phi' + \left[2\left(\frac{a'}{a}\right)' + \left(\frac{a'}{a}\right)^2\right]\Phi = 4\pi G a^2\left[\delta p_x + \frac{\gamma - 1}{2}\left(\delta\rho_m + \delta\rho_x - 3\delta p_x\right)\right]\;.
\end{eqnarray}
For $p_x = -\rho_x$, Eqs.~(\ref{contpertx}) and (\ref{eulerpertx}) become
\begin{equation}
 \delta\rho_x' = \frac{\gamma - 1}{2}\left(\delta\rho_m' + 4\delta\rho_x'\right)\;, \qquad \partial^i\delta \rho_x = \frac{\gamma - 1}{2}\partial^i\left(\delta\rho_m + 4\delta\rho_x\right)\;,
\end{equation}
and therefore we have
\begin{equation}\label{deltaxdeltaeq}
 \delta \rho_x = \frac{\gamma - 1}{2(3 - 2\gamma)}\delta\rho_m\;,
\end{equation}
i.e. the perturbation in the $x$ fluid is proportional to the one for matter, the same relation \eqref{fpr1} we found using the synchronous coordinate condition. Combining Eqs. (\ref{E00}) and (\ref{Eij}) with Eq. (\ref{deltaxdeltaeq}) we obtain
 \begin{eqnarray}\label{E00choice}
 \Delta\Phi - 3\frac{a'}{a}\left(\frac{a'}{a}\Phi + \Phi'\right) = 4\pi G a^2\delta\rho_m\;,\\
\label{Eijchoice} \Phi'' + 3\frac{a'}{a}\Phi' + \left[2\left(\frac{a'}{a}\right)' + \left(\frac{a'}{a}\right)^2\right]\Phi = 0\;.
\end{eqnarray}
These equations are identical to the corresponding ones for the $\Lambda$CDM model (note that the background is also the same). In particular, the first one confirms that matter perturbations are not affected by the $x$ fluid (with $\omega_x = -1$), even if the latter also agglomerates.

\section{Non-linear Regime}

We follow \cite{Malik:2008im} for the treatment of the second order regime of cosmological scalar perturbations. In particular, consider Eqs.~(4.15)--(4.17) of \cite{Malik:2008im} for the $x$ component of our model in the Newtonian gauge (i.e. zero metric coefficient $B$), with zero anisotropic pressure and $p_x = - \rho_x$, $\delta p_x = -\delta\rho_x$ (we assume this relation at all orders):
\begin{equation}\label{TxEq}
^{(2)}{T^0_x}_0 = \delta\rho_x^{(2)}\;,\qquad ^{(2)}{T^0_x}_i = 0\;, \qquad ^{(2)}{T^i_x}_0 = 0\;, \qquad ^{(2)}{T^i_x}_j = -\delta p_x^{(2)}\delta^i_j\;,
\end{equation}
where the subscript $(\cdot)$ refers to the perturbation order. Moreover, the second-order trace has the following form:
\begin{equation}
^{(2)}T_x = ^{(2)}{T^0_x}_0 + ^{(2)}{T^k_x}_k = \delta\rho_x^{(2)} - 3\delta p^{(2)}_x  = 4\delta\rho_x^{(2)}\;.
\end{equation}
On the other hand, for the matter component we have $p_m = 0$. Hence Eqs.~(4.15), (4.17) of \cite{Malik:2008im} become
\begin{equation}
^{(2)}{T^0_m}_0 = \delta\rho_x^{(2)} + 2\rho_mv_k^{(1)} v^{(1)k}\;, \qquad ^{(2)}{T^i_m}_j = -2\rho_mv_k^{(1)} v^{(1)k}\;,
\end{equation}
where the $v_k^{(1)}$ represents the velocity perturbation at first order. Hence, for the second-order trace we get
\begin{equation}
^{(2)}T_m = { ^{(2)}{T^0_m}_0} + { ^{(2)}{T^k_m}_k} = \delta\rho_m^{(2)}.
\end{equation}
Now consider the equation
\begin{equation}
{{T^\mu_x}_\nu}_{;\mu} = \frac{\gamma - 1}{2}(T_x + T_m)_{,\nu}\;,
\end{equation}
at second order:
\begin{eqnarray}
\partial_\mu\,{ ^{(2)}{{T^\mu_x}_\nu}} &+& { ^{(0)}\Gamma^\mu_{\mu\rho}} { ^{(2)}{{T^\rho_x}_\nu}} + { ^{(1)}\Gamma^\mu_{\mu\rho}}{ ^{(1)}{{T^\rho_x}_\nu}} +{ ^{(2)}\Gamma^\mu_{\mu\rho}} { ^{(0)}{{T^\rho_x}_\nu}} \nonumber\\
&-& { ^{(0)}\Gamma^\rho_{\mu\nu}}{ ^{(2)}{{T^\mu_x}_\rho}} - { ^{(1)}\Gamma^\rho_{\mu\nu}}{ ^{(1)}{{T^\mu_x}_\rho}} - { ^{(2)}\Gamma^\rho_{\mu\nu}}{ ^{(0)}{{T^\mu_x}_\rho}} = \frac{\gamma - 1}{2}\left[{ ^{(2)} T_m} + { ^{(2)} T_x}\right]_{_,\nu}\;.
\end{eqnarray}
Making $\nu = 0$, and taking into account the non-vanishing values of the perturbations, we obtain:
\begin{eqnarray}
\partial_0\,{ ^{(2)}{{T^0_x}_0}} &+& { ^{(0)}\Gamma^\mu_{\mu0}} { ^{(2)}{{T^0_x}_0}} + { ^{(1)}\Gamma^\mu_{\mu0}}{ ^{(1)}{{T^0_x}_0}} +{ ^{(2)}\Gamma^\mu_{\mu0}} { ^{(0)}{{T^0_x}_0}} \nonumber\\
&-& { ^{(0)}\Gamma^\rho_{\mu0}}{ ^{(2)}{{T^\mu_x}_\rho}} - { ^{(1)}\Gamma^\rho_{\mu_0}}{ ^{(1)}{{T^\mu_x}_\rho}} - { ^{(2)}\Gamma^\rho_{\mu0}}{ ^{(0)}{{T^\mu_x}_\rho}} = \frac{\gamma - 1}{2}\left[{ ^{(2)} T_m} + { ^{(2)} T_x}\right]_{_,0}\;.
\label{fund}
\end{eqnarray}
On the other hand:
\begin{eqnarray}
{ ^{(0)}\Gamma^\rho_{\mu0}}{ ^{(2)}{{T^\mu_x}_\rho}} &=& { ^{(0)}\Gamma^0_{00}}{ ^{(2)}{{T^0_x}_0}} + { ^{(0)}\Gamma^i_{j0}}{ ^{(2)}{{T^j_x}_i}}\nonumber\\
&=& \frac{a'}{a}\left[^{(2)}{{T^0_x}_0} + ^{(2)}{{T^k_x}_k}\right] = 4\frac{a'}{a}\delta\rho_x^{(2)}\;,
\end{eqnarray}
and
\begin{equation}
{ ^{(0)}\Gamma^\mu_{\mu0}} { ^{(2)}{{T^0_x}_0}} = 4\frac{a'}{a}\delta\rho_x^{(2)}\;.
\end{equation}
Hence, the second term in the first line of Eq.~(\ref{fund}), cancels with the first term of the second line. In the same way, the third term and fourth term of the first line cancel with the second and third of the second line. We end up with
\begin{equation}\label{deltaxdeltaeq2ord}
{\delta\rho_x^{(2)}}' = \frac{\gamma - 1}{2}\left[{4\delta\rho_x^{(2)}}' + {\delta\rho_m^{(2)}}'\right]\;, \qquad \Rightarrow \qquad \delta \rho^{(2)}_x = \frac{\gamma - 1}{2(3 - 2\gamma)}\delta\rho^{(2)}_m\;,
\end{equation}
and we recover, up to an integration constant, Eq.~\eqref{deltaxdeltaeq}, now at the second order. Therefore, it seems that in the Rastall cosmological models here investigated, the evolution of perturbations in the $x$ fluid mimic those in the matter up to the second-order perturbation regime.

The left-hand side of Eq.~\eqref{eq1R} is the same as in Einstein's theory, for the background and for the perturbations evolution. 
% So, let us focus on the right-hand side which we rename, for simplicity,
% \begin{equation}
% S^\mu{}_\nu = T^\mu{}_\nu - \frac{(\gamma - 1)}{2}\delta^\mu{}_\nu T\;,
% \end{equation}
% in mixed components. At the first order, using Eq.~\eqref{deltaxdeltaeq} and $\delta p_x = -\delta\rho_x$, we find
% \begin{equation}
% ^{(1)}S^0{}_0 = -\delta\rho^{(1)}_m - \delta\rho^{(1)}_x - \frac{\gamma - 1}{2}\left[-\delta\rho^{(1)}_m - \delta\rho^{(1)}_x + 3\delta p^{(1)}_x\right] = -\delta\rho^{(1)}_m\;,
% \end{equation}
% and, following the same strategy, we also have
% \begin{equation}
% { ^{(1)}}S^0{}_i = \rho_m v^{(1)}_i\;, \qquad { ^{(1)}}S^i{}_j = -(3 - 2\gamma)\delta\rho^{(1)}_x \delta^i{}_j\;.
% \end{equation}
% The last equation carries a dependence on $\gamma$. This is necessary in order the Bianchi's identities to be consistent with (\ref{deltaxdeltaeq}). Let us turn to the second order level. 
As for the right hand side, using Eq.~(4.16) of \cite{Malik:2008im} for matter ($p_m = 0$), i.e.
\begin{equation}
{^{(2)}}{T^0_m}{}_j = -\rho_mv^{(2)}_j + 4\psi^{(1)}v^{(1)}_j + 2\phi^{(1)}v^{(1)}_j\;,
\end{equation}
we find 
\begin{equation}
{}^{(2)}T^0{}_0 - \frac{(\gamma - 1)}{2}{}^{(2)}T = \delta\rho^{(2)}_m + 2\rho_m v_k^{(1)}v^{(1)k}\;.
\end{equation}
This is the same relation that one can find in the usual $\Lambda$CDM model. Following the same steps, we have
\begin{eqnarray}
{}^{(2)}T^0{}_i &=& { ^{(2)}}{T^0_m}{}_i = -\rho_m\left[v^{(2)}_i - 4\psi^{(1)}v^{(1)}_i - 2\phi^{(1)}v^{(1)}_i\right] - 2\delta\rho^{(1)}_m v^{(1)}_i\;,\\
{}^{(2)}T^i{}_j - \frac{(\gamma - 1)}{2}\delta^i{}_j {}^{(2)}T &=& -2\rho_m v^{(1)i} v^{(1)}_j\;.
\end{eqnarray}
All the relations are the same as in the $\Lambda$CDM model. Therefore, even at the second order, considering scalar perturbations only, the evolution of matter perturbations is not affected by the agglomeration of the $x$ component.

However, in the full non-linear regime one should expect some differences between Rastall's model and the $\Lambda$CDM one. This because we have now two components agglomerating and they would gravitationally interact affecting in some manner, hopefully detectable, the growth of structure. Such analysis is beyond the scope of the present paper, nevertheless we can argue some hints by considering the simple case of a spherical top-hat collapse. Let us follow the growth of an inhomogeneity in the past, when the universe behaved as an Einstein-de Sitter one since the effects of a cosmological constant were safely negligible. Following \cite{coles1995cosmology}, the growth of a spherical homogeneous region is described by the equation
\begin{equation}\label{sphercoll}
 \left(\frac{\dot a}{a_i}\right)^2 = H_i^2\left[\Omega_p(t_i)\frac{a_i}{a} + 1 - \Omega_p(t_i)\right]\:,
\end{equation}
where $a_i$ is the scale factor computed at some time $t_i$ at which the collapse begins and $\Omega_p = 1 + \delta$ is the density parameter of the collapsing region, determined by the density contrast of the fluid there contained. Here comes the relevant difference. For the $\Lambda$CDM model, the only collapsing component is matter and, therefore, $\Omega_p = 1 + \delta_m$. On the other hand, in the Rastall models we presented, an unknown fluid $x$ with a vacuum equation of state also can agglomerate. For this reason, the density parameter of the collapsing region assumes now the form
\begin{equation}
 \Omega_p = 1 + \frac{\delta\rho_m}{\rho_m + \rho_x} + \frac{\delta\rho_x}{\rho_m + \rho_x}\;.
\end{equation}
Considering now Eq.~\eqref{xevoeta} with $\eta_m = \eta_x = 0$, and neglecting the contribution of the constant $\bar\rho$, we can write
\begin{equation}
 \Omega_p = 1 + \frac{2(3 - 2\gamma)}{5 - 3\gamma}\delta_m + \frac{\gamma - 1}{5 - 3\gamma}\delta_x\;.
\end{equation}
Equation~\eqref{sphercoll} is assumed to hold true for both the models and it is easy to solve, giving the size of the perturbation, once we know $\Omega_p(t_i)$. For the $\Lambda$CDM we have that $\Omega_p(t_i) = 1 + \delta_{m}(t_i)$, whereas in the Rastall-type cosmological model we have one degree of freedom more than the $\Lambda$CDM one, because of the initial condition $\delta_x(t_i)$. So, in principle, one can construct the same collapsing history as in the $\Lambda$CDM one but with a different amount of matter, depending on the value of $\gamma$.

\section{Discussion and Conclusions}

We have shown that there is a subset of cosmological scenarios based on Rastall's energy-momentum tensor non-conservation that are equivalent to the $\Lambda$CDM cosmology, except for one aspect: vacuum energy may agglomerate. Two effects could allow to discriminate between the two models: non-linear effects in the matter power spectrum and the transfer function for cosmological perturbations. Vacuum energy is negligible in the past, hence the impact of its fluctuation in the evolution of the perturbations in the primordial periods of the evolution of the universe may not be relevant. However, according to the scenario described above, dark energy must be present in virialized systems, like galaxies and cluster of galaxies, and the effect of agglomeration of dark energy must be relevant at this level. Therefore, Rastall's cosmology and the $\Lambda$CDM model seem to be distinguishable only at the non-linear regime of the evolution of cosmic perturbations. Though we have given some hints of the latter possibility addressing the simple case of a spherical top-hat collapse, a deeper analysis shall be the subject of future investigation.

\subsection*{Acknowledgements}

The authors thank the anonymous referee for his kind and useful suggestions. J. C. F. thanks CNPq (Brasil) for partial financial support.


\begin{thebibliography}{99}
\bibitem{Cartan1922}
\'E.~Cartan,
``Sur une g\'en\'eralisation de la notion de courbure de Riemann et les espaces \`a torsion,'' 
Acad. Sci. Paris, Comptes Rend. {\bf 174} (1922) 593-595
\bibitem{Cartan1923}
\'E.~Cartan, 
``Sur les vari\'et\'es \`a connexion affine et la th\'eorie de la relativit\'e g\'en\'eralis\'ee,'' 
Annales Scientifiques de l'\'Ecole Normale Superieure S\'er. 3, {\bf 40} (1923) 325-412
\bibitem{Brans1961}
C.~Brans and R.~H.~Dicke,
``Mach's principle and a relativistic theory of gravitation,'' 
Phys. Rev. {\bf 124} (1961) 925-935
\bibitem{Rosen1971}
N.~Rosen,
``Theory of gravitation,'' 
Physical Review D {\bf 3} (1971) 2317
\bibitem{Rastall1973}
  P.~Rastall,
  ``Generalization of the einstein theory,''
  Phys.\ Rev.\ D\ {\bf 6} (1972) 3357.
\bibitem{Rastall1976}
  P.~Rastall,
  ``A Theory of Gravity,''
  Can.\ J.\ Phys.\ \ {\bf 54} (1976) 66.
\bibitem{Moffat:1994hv}
  J.~W.~Moffat,
  ``Nonsymmetric gravitational theory,''
  Phys.\ Lett.\ B {\bf 355} (1995) 447
  [gr-qc/9411006].
\bibitem{Bekenstein:2004ne}
  J.~D.~Bekenstein,
  ``Relativistic gravitation theory for the MOND paradigm,''
  Phys.\ Rev.\ D {\bf 70} (2004) 083509
   [Erratum-ibid.\ D {\bf 71} (2005) 069901]
  [astro-ph/0403694].
\bibitem{Bondi1948} 
H.~Bondi and T.~Gold,
``The Steady-State Theory of the Expanding Universe,''
Monthly Notices of the Royal Astronomical Society {\bf 108} (1948) 252
\bibitem{Hoyle1948} 
F.~Hoyle,
``A new model for the expanding universe,''
Monthly Notices of the Royal Astronomical Society {\bf 108} (1948) 372
\bibitem{Jordan:1949zz}
  P.~Jordan,
  ``Formation of the Stars and Development of the Universe,''
  Nature {\bf 164} (1949) 637.
\bibitem{Gibbons:1977mu}
  G.~W.~Gibbons and S.~W.~Hawking,
  ``Cosmological Event Horizons, Thermodynamics, and Particle Creation,''
  Phys.\ Rev.\ D {\bf 15} (1977) 2738.
\bibitem{Parker:1971pt}
  L.~Parker,
  ``Quantized fields and particle creation in expanding universes. 2.,''
  Phys.\ Rev.\ D {\bf 3} (1971) 346
   [Erratum-ibid.\ D {\bf 3} (1971) 2546].
\bibitem{Ford:1986sy}
  L.~H.~Ford,
  ``Gravitational Particle Creation and Inflation,''
  Phys.\ Rev.\ D {\bf 35} (1987) 2955.
\bibitem{Smalley1984}
L.~L.~Smalley,
``Variational Principle for a Prototype Rastall Theory of Gravitation,''
Nuovo Cim.\ B {\bf 80} 1 (1984) 42
\bibitem{Batista:2010nq}
  C.~E.~M.~Batista, J.~C.~Fabris and M.~H.~Daouda,
  ``Testing the Rastall's theory using matter power spectrum,''
  Nuovo Cim.\ B {\bf 125} (2010) 957
  [arXiv:1004.4603 [astro-ph.CO]].
\bibitem{Fabris:2011rm}
  J.~C.~Fabris, T.~C.~C.~Guio, M.~Hamani Daouda and O.~F.~Piattella,
  ``Scalar models for the generalized Chaplygin gas and the structure formation constraints,''
  Grav.\ Cosmol.\  {\bf 17} (2011) 259
  [arXiv:1011.0286 [astro-ph.CO]].
\bibitem{Fabris:2011wz}
  J.~C.~Fabris, M.~H.~Daouda and O.~F.~Piattella,
  ``Note on the Evolution of the Gravitational Potential in Rastall Scalar Field Theories,''
  arXiv:1109.2096 [astro-ph.CO].
\bibitem{Capone:2009xm}
  M.~Capone, V.~F.~Cardone and M.~L.~Ruggiero,
  ``Accelerating cosmology in Rastall's theory,''
  Nuovo Cim.\ B {\bf 125} (2011) 1133
  [arXiv:0906.4139 [astro-ph.CO]].
\bibitem{Amanullah:2010vv}
  R.~Amanullah, C.~Lidman, D.~Rubin, G.~Aldering, P.~Astier, K.~Barbary, M.~S.~Burns and A.~Conley {\it et al.},
  ``Spectra and Light Curves of Six Type Ia Supernovae at 0.511 < z < 1.12 and the Union2 Compilation,''
  Astrophys.\ J.\  {\bf 716} (2010) 712
  [arXiv:1004.1711 [astro-ph.CO]].
\bibitem{Gaztanaga:2008xz}
  E.~Gaztanaga, A.~Cabre and L.~Hui,
  ``Clustering of Luminous Red Galaxies IV: Baryon Acoustic Peak in the Line-of-Sight Direction and a Direct Measurement of H(z),''
  Mon.\ Not.\ Roy.\ Astron.\ Soc.\  {\bf 399} (2009) 1663
  [arXiv:0807.3551 [astro-ph]].
\bibitem{Stern:2009ep}
  D.~Stern, R.~Jimenez, L.~Verde, M.~Kamionkowski and S.~A.~Stanford,
  ``Cosmic Chronometers: Constraining the Equation of State of Dark Energy. I: H(z) Measurements,''
  JCAP {\bf 1002} (2010) 008
  [arXiv:0907.3149 [astro-ph.CO]].
\bibitem{Komatsu:2010fb}
  E.~Komatsu {\it et al.}  [WMAP Collaboration],
  ``Seven-Year Wilkinson Microwave Anisotropy Probe (WMAP) Observations: Cosmological Interpretation,''
  Astrophys.\ J.\ Suppl.\  {\bf 192} (2011) 18
  [arXiv:1001.4538 [astro-ph.CO]].
\bibitem{Eisenstein:2005su}
  D.~J.~Eisenstein {\it et al.}  [SDSS Collaboration],
  ``Detection of the baryon acoustic peak in the large-scale correlation function of SDSS luminous red galaxies,''
  Astrophys.\ J.\  {\bf 633} (2005) 560
  [astro-ph/0501171].
\bibitem{Fabris:1998hr}
  J.~C.~Fabris, R.~Kerner and J.~Tossa,
  ``Perturbative analysis of generalized Einstein's theories,''
  Int.\ J.\ Mod.\ Phys.\ D {\bf 9} (2000) 111
  [gr-qc/9806059].
\bibitem{Padmanabhan1993}
  T.~Padmanabhan,
  ``Structure formation in the Universe,''
  Cambridge University Press (1993)
\bibitem{Ma:1995ey}
  C.~-P.~Ma and E.~Bertschinger,
  ``Cosmological perturbation theory in the synchronous and conformal Newtonian gauges,''
  Astrophys.\ J.\  {\bf 455} (1995) 7
  [astro-ph/9506072].
\bibitem{Sachs1967}
  R.~K.~Sachs and A.~M.~Wolfe,
  ``Perturbations of a cosmological model and angular variations of the microwave background,''
  Astrophys.\ J.\ \ {\bf 147} (1967) 73
   [Gen.\ Rel.\ Grav.\ \ {\bf 39} (2007) 1929].
\bibitem{Mukhanov2005}
  V.~Mukhanov,
  ``Physical Foundations of Cosmology,''
  Cambridge University Press (2005)
\bibitem{Malik:2008im}
  K.~A.~Malik and D.~Wands,
  ``Cosmological perturbations,''
  Phys.\ Rept.\  {\bf 475} (2009) 1
  [arXiv:0809.4944 [astro-ph]].
\bibitem{coles1995cosmology}
  P.~Coles and F.~ Lucchin,
  ``Cosmology. The origin and evolution of cosmic structure,''
  Chichester: Wiley, 2nd edition (2002)
\end{thebibliography}
\end{document}